\documentclass[12pt]{article}
\usepackage{a4}
\usepackage{subfigure}
\usepackage{graphicx}

\begin{document}
\title{Convergence of the homotopy analysis method}
\author{M. Turkyilmazoglu}
\maketitle
\begin {center}
{Mathematics Department, University of Hacettepe, 06532-Beytepe,
Ankara, Turkey, Tel: 009003122977850, Fax: 009003122972026, Email:
turkyilm@hotmail.com}
\end {center}

\renewcommand{\baselinestretch}{1.}
\begin{abstract}
The homotopy analysis method is studied in the present paper. The
question of convergence of the homotopy analysis method is
resolved. It is proven that under a special constraint the
homotopy analysis method does converge to the exact solution of
the sought solution of nonlinear ordinary or partial differential
equations. An optimal value of the convergence control parameter
is given through the square residual error. An error estimate is
also provided. Examples, including the Blasius flow, clearly
demonstrate why and on what interval the corresponding homotopy
series generated by the homotopy analysis method will converge to
the exact solution.
\end{abstract}
{\bf Key words:} Nonlinear equations, Approximate solution,
Homotopy analysis method, Convergence

\section{Introduction}
\label{introduction} The search for a better and easy to use tool
for the solution of nonlinear equations illuminating the nonlinear
phenomena of our life keeps continuing.

A variety of methods therefore were proposed to find approximate
solutions. One of the most recent popular technique is the
homotopy analysis method, which is a combination of the classical
perturbation technique and homotopy concept as used in topology.
In the homotopy analysis method, which requires neither a small
parameter nor a linear term in a differential equation, a homotopy
with an embedding parameter $p \in [0, 1]$ is constructed. In
\cite{Liao92a} a basic idea of homotopy analysis method for
solving nonlinear differential equations was presented. In this
method, the solution is considered as the sum of an infinite
series, which converges rapidly to accurate solutions. A numerous
nonlinear problems in science, finance and engineering were
recently treated by the method, see at least
\cite{Abbasbandy2006}, \cite{Zhu2006}, \cite{Hayat2008},
\cite{VanGorder2008}, \cite{Akyildiz2008}, \cite{Wu2009},
\cite{Liang2009},\cite{Molabahrami2009} and \cite{Zhao2009}.
Particularly, a few new solutions of some nonlinear problems were
found by means of the method \cite{Liao2007a}, which were
neglected by other analytic methods and even by numerical
techniques. All of these show the potential of the homotopy
analysis method for strongly nonlinear problems. Recently an
optimal homotopy analysis approach for strongly nonlinear
differential equations were proposed in \cite{Liao2010}. Even
though a great deal of equations were solved using the homotopy
analysis method, the question of convergence of the method is yet
to be answered.

We in the present paper investigate the homotopy analysis
technique from a mathematical point of view. The aim is to analyze
the method and to show that under certain circumstances the
homotopy analysis method converges to the exact solution desired,
without a prior knowledge of the exact solution. An optimal value
of the convergence control parameter is defined through the square
residual error concept. Our another emphasis is to address the
error estimate of the approximate solution. The given theorem is
justified exemplifying it by basic examples from ordinary and
partial differential equations from the literature. The presented
theory not only gives the convergence, but it also provides the
information about the interval of convergence of the homotopy
series.

In the rest of the paper, \S \ref{Method} lays the basis of
homotopy analysis method. A theory is outlined in \S \ref{Results}
for the convergence and error estimate. Illustrative examples in
\S \ref{Examp} are followed by the conclusions in \S
\ref{conclusions}.

\section{The Homotopy Analysis Method}
\label{Method} Liao \cite{Liao92a} described the early form of the
homotopy analysis method in 1992. The essential idea of this
method is to introduce a homotopy parameter, say $p$, which varies
from 0 to 1 and a nonzero auxiliary parameter so-called the
convergence control parameter $h$. At $p=0$, the system of
equations usually has been reduced to a simplified form which
normally admits a rather simple solution. As $p$ gradually
increases continuously toward 1, the system goes through a
sequence of deformations, and the solution at each stage is close
to that at the previous stage of the deformation. Eventually at
$p=1$, the system takes the original form of the equation and the
final stage of the deformation gives the desired solution. To
illustrate the basic ideas of this method, consider the nonlinear
boundary value problem
\begin{eqnarray}\label{eq1}
   N(u(r))=0;\;r\in \Omega, \quad B(u(r),{du\over dn}) = 0;\;r\in
   \Gamma
\end{eqnarray}
where $u(r)$ defined over the region $\Omega$ is the function to
be solved under the boundary constraints in $B$ defined over the
boundary $\Gamma$ of $\Omega$. The homotopy analysis technique
defines a homotopy $u(r,p): R\times [0,1]\rightarrow R$ so that
\begin{eqnarray}\label{eq2}
  H(u,p) &=& (1-p)[L(u)-L(u_0)]+p h N(u),
\end{eqnarray}where $L$ is a suitable auxiliary linear operator,
$u_0$ is an initial approximation of equation (\ref{eq1})
satisfying exactly the boundary conditions. It is obvious from
equation (\ref{eq2}) that
\begin{eqnarray}\label{eq3}
  H(u,0) &=& L(u)-L(u_0), \quad H(u,1)\; =\; N(u).
\end{eqnarray}As $p$ moves from 0 to 1, $u(r,p)$ moves from
$u_0(r)$ to $u(r)$. In topology, this called a deformation and
$L(u)-L(u_0)$ and $N(u)$ are said to be homotopic. Our basic
assumption is that the solution of equation (\ref{eq2}) when
equated to zero can be expressed as a power series in $p$
\begin{eqnarray}\label{eq4}
   u(r,p) &=& u_0(r)+pu_1(r)+p^2u_2(r)+\cdots =
   \sum_{k=0}^\infty u_k(r)p^k.
\end{eqnarray}
The appropriate solutions of the coefficients $u_k(r)$ in
(\ref{eq4}) can be found from the homotopy deformation equations,
see \cite{Liao2003book}. Hence, the approximate solution of
equation (\ref{eq1}) can be readily obtained as
\begin{eqnarray}\label{eq5}
   u(r)=\lim_{p\to 1}u(r,p) = \sum_{k=0}^\infty u_k(r).
\end{eqnarray}It was found that the auxiliary parameter $h$ can
adjust and control the convergence region and rate of homotopy
series solutions (\ref{eq4}). Whenever the series (\ref{eq4}) is
known to be convergent, then (\ref{eq5}) represents the exact
solution of (\ref{eq1}), as proved in \cite{Liao2003book}.

\section{A convergence Theorem and error estimate}\label{Results}
Using the methodology underlined above, the number of problems
treated by the homotopy analysis method approaches a couple of
hundreds now. However, the very basic question of why the series
obtained by setting $p=1$ in (\ref{eq4}) should be convergent
remains unanswered till today. To remedy this issue up to a point,
we provide the subsequent theorems here. It should be noted that
how to find a proper convergence control parameter $h$, or even
better, to get a faster convergent one, to be used in Theorem 1
will be clarified later in this section.

\vspace{0.2in}{\bf Theorem 1.} Suppose that $A\subset R$ be a
Banach space donated with a suitable norm $\|\|$ (depending on the
physical problem considered), over which the sequence $u_k(t)$ of
(\ref{eq4}) is defined for a prescribed value of $h$. Assume also
that the initial approximation $u_0(t)$ remains inside the ball of
the solution $u(t)$. Taking $r\in R$ be a constant, the following
statements hold true:

{\bf (i)} If $\|v_{k+1}(t)\|\le r \|v_k(t)\|$ for all $k$, given
some $0<r<1$, then the series solution $u(t,p) = \sum_{k=0}^\infty
u_k(t)p^k$, defined in (\ref{eq4}) converges absolutely at $p=1$
to (\ref{eq5}) over the domain of definition of $t$,

{\bf (ii)} If $\|v_{k+1}(t)\|\ge r \|v_k(t)\|$ for all $k$, given
some $r>1$, then the series solution $u(t,p) = \sum_{k=0}^\infty
u_k(t)p^k$, defined in (\ref{eq4}) diverges at $p=1$ over the
domain of definition of $t$.

\vspace{0.2in}{\bf Proof.} {\bf (i)} In compliance with the ratio
test for the power series in $p$, the proof is clear. However, in
order to give an estimate to the truncation error of homotopy
analysis method, we shortly give the whole proof here. If $S_n(t)$
denote the sequence of partial sum of the series (\ref{eq5}), we
need to show that $S_n(t)$ is a Cauchy sequence in $A$. For this
purpose, consider,
\begin{eqnarray}\label{eq6}
  \|S_{n+1}(t)-S_n(t)\| = \|u_{n+1}(t)\|\le r \|u_n(t)
  \|\le r^2 \|u_{n-1}(t)\|\le \cdots \le r^{n+1}\|u_0(t)\|.
\end{eqnarray}
It should be remarked that owing to (\ref{eq6}), all the
approximations produced by the homotopy method (\ref{eq2}) will
lie within the ball of $u(t)$. For every $m,n \in N$, $n\ge m$,
making use of (\ref{eq6}) and the triangle inequality
successively, we have,
\begin{eqnarray}\label{eq7}
  \|S_{n}(t)-S_m(t)\| = \|(S_n(t)-S_{n-1}(t))+(S_{n-1}(t)-S_{n-2}(t))+
  \cdots+(S_{m+1}(t)-S_m(t))\|\\\nonumber \le {1-r^{n-m}\over
  1-r}r^{m+1}\|u_0(t)\|.
\end{eqnarray}
Since $0<r<1$, we get from (\ref{eq7})
\begin{eqnarray}\label{eq8}
  \lim_{n,m\to\infty}\|S_{n}(t)-S_m(t)\| &=& 0.
\end{eqnarray}
Therefore, $S_n(t)$ is a Cauchy sequence in the Banach space $A$,
and this implies that the series solution (\ref{eq5}) is
convergent. This completes the proof {\bf (i)}. $\diamond$

The proof of {\bf (ii)} follows from the fact that under the
hypothesis supplied in {\bf (ii)}, there exist a number $l$,
$l>r>1$, so that the interval of convergence of the power series
(\ref{eq4}) is $|p|<1/l<1$, which obviously excludes the case of
$p=1$. $\diamond$

\vspace{0.2in}{\bf Remark 1.} Since the finite number of terms
does not affect the convergence, Theorem 1 is equally valid if the
inequalities stated in {\bf (i-ii)} are true for sufficiently
large $k'$s. Thus, it is sufficient to keep track of magnitudes of
the ratios $rat_k$ defined by
\begin{eqnarray}\label{rat}
  rat_k &=& {\|v_{k+1}(t)\|\over \|v_k(t)\|},
\end{eqnarray}and whether they remain less than unity.

\vspace{0.2in}{\bf Remark 2.} By enforcing the ratio in {\bf (i)}
to hold true in the infinite limit, the validity region of $t$ for
the series solution can also be constructed, see Example 1 below.

\vspace{0.2in}{\bf Theorem 2.} If the series solution defined in
(\ref{eq5}) is convergent, then it converges to an exact solution
of the nonlinear problem (\ref{eq1}).

\vspace{0.2in}{\bf Proof.} The proof can be found in
\cite{Liao2003book}.

\vspace{0.2in}{\bf Theorem 3.} Assume that the series solution
$\sum_{n=0}^\infty u_n(t)$, defined in (\ref{eq5}) is convergent
to the solution $u(t)$ for a prescribed value of $h$. If the
truncated series $\sum_{n=0}^Mu_n(t)$ is used as an approximation
to the solution $u(t)$ of problem (\ref{eq1}), then an upper bound
for the error, $E_M(t)$, is estimated as
\begin{eqnarray}\label{eq10}
  E_M(t) \le {r^{M+1}\over 1-r}\|u_0(t)\|.
\end{eqnarray}

\vspace{0.2in}{\bf Proof.} Making use of the inequality
(\ref{eq7}) of Theorem 1, we immediately obtain
\begin{eqnarray}\label{eq11}
  \|u(t)-S_M(t)\| \le {1-r^{n-M}\over
  1-r}r^{M+1}\|u_0(t)\|,
\end{eqnarray}
and taking into account $(1-r^{n-M})<1$, (\ref{eq11}) leads to the
desired formula (\ref{eq10}). This completes the proof. $\diamond$

\vspace{0.2in}{\bf Remark 3.} An optimal value of the convergence
control parameter $h$ can be found by means of the exact square
residual error integrated in the whole region of interest
$\Gamma$, at the order of approximation $M$, that is,
\begin{eqnarray}\label{eqres1}
  Res(h) &=& \int_\Gamma [N(\sum_{k=0}^Mu_k(r))]^2dr.
\end{eqnarray}
Obviously, the more quickly $Res(h)$ in (\ref{eqres1}) decreases
to zero, the faster the corresponding homotopy series solution
(\ref{eq5}) converges. So, at the given order of approximation
$M$, the corresponding optimal value of the convergence control
parameter $h$ is given by the minimum of $Res(h)$, corresponding
to a nonlinear algebraic equation of the form
\begin{eqnarray}\label{eqres2}
  {d Res \over d h} &=& 0.
\end{eqnarray}

Therefore, the convergence control parameter obtained via
(\ref{eqres1}-\ref{eqres2}) can be supplied into the Theorem 1.
However, it is unfortunate that the exact square residual error
$Res(h)$ defined by (\ref{eqres1}) needs too much CPU time to
calculate even if the order of approximation is not very high, and
thus is often useless in practice. To avoid the time-consuming
computation, Liao in \cite{Liao2003book} suggested to investigate
the convergence of some special quantities, which often have
important physical meanings. For example, one can consider the
convergence of $u'(0)$ and $u''(0)$ of a nonlinear differential
equation (\ref{eq1}), if they are unknown. It is found by the
homotopy analysis researchers that there often exists such a
region that certain values of $h$ give a convergent series
solution of such kind of quantities. Besides, such a region can be
found, although approximately, by plotting the curves of these
unknown quantities versus $h$. These curves are called {\em
$h$-curves} or {\em curves for convergence-control parameter},
which have been successfully applied in many nonlinear problems as
cited herein. This approach constitutes another way of finding
proper values of $h$ for the Theorem 1.

\vspace{0.2in}{\bf Remark 4.} Similar to the constant $h-$curves
idea of Liao, an approximate interval of convergence for $h$ can
be determined by application of Theorem 1 to some certain physical
quantities, see Example 3 below.

\section{Illustrative Examples}\label{Examp}
To illustrate the validity of the Theorems outlined, we take into
account the following examples taken from the homotopy analysis
studies in the literature.

\vspace{0.2in}{\bf Example 1.} Consider the first-order
nonlinear differential equation \cite{Liao2003book}
\begin{eqnarray}\label{eq12}
   u'+u^2 &=& 1,\quad u(0) = 0,
\end{eqnarray}that governs the steady free convection flow over
a vertical semi-infinite flat plate which is embedded in a fluid
saturated porous medium of ambient temperature \cite{Crane70} and
also the steady-state boundary-layer flows over a permeable
stretching sheet \cite{Keller2000}. In accordance with Theorem 1,
choosing $u_0(t)=1-e^{-2t}$ and $L={d\over dt}+2$, the homotopy
series solution via the homotopy approach (\ref{eq2}) can be
straightforwardly constructed. Employing the $L^2$ norm in $R$,
then Theorem 1 assures the convergency of (\ref{eq12}) provided
that
\begin{eqnarray}\label{eq13}
  {\|u_{n+1}(t)\| \over \| u_n(t)\|} <1.
\end{eqnarray}
Table \ref{Table 1} presents evolution of the ratio (\ref{eq13})
for a variety of the convergence control parameter $h$ for the
problem (\ref{eq12}). Additionally, Table \ref{Table 1} gives the
error, see the last column, defined by
\begin{eqnarray}\label{eq14}
  err &=& \int_0^\infty|u_e(t)-u(t)|dt,
\end{eqnarray}where $u_e(t)$ is the solution of (\ref{eq12})
calculated numerically. Data displayed in Table \ref{Table 1}
clearly explains why the homotopy analysis method generates
completely convergent series solution to the problem (\ref{eq12})
for the chosen parameters $h$.
\begin{table}[!htb]
\begin{center}
\begin{tabular}{|c|c|c|c|c|c|c|c|}
  \hline $h$ &$M=1$    & $M=10$  & $M=20$  & $M=30$  & $M=40$  & $M=50$  & $err$\\\hline
  2.0        & 0.44721 & 0.87228 & 0.93102 & 0.95275 & 0.96407 & 0.97044 & $9.609\times 10^{-5}$ \\
  1.5        & 0.15811 & 0.43509 & 0.46523 & 0.47625 & 0.48196 & 0.48517 & $3.768\times 10^{-20}$\\
  1.3        & 0.15811 & 0.31877 & 0.33354 & 0.33881 & 0.34152 & 0.34304 & $8.261\times 10^{-26}$\\
  1.0        & 0.31623 & 0.45743 & 0.47701 & 0.48425 & 0.48802 & 0.49015 & $8.384\times 10^{-18}$\\
  0.5        & 0.65192 & 0.69906 & 0.71905 & 0.72798 & 0.73295 & 0.73583 & $1.492\times 10^{-8}$ \\\hline
\end{tabular}
\end{center}
\caption{The evolution of the ratio (\ref{eq13}) for the equation
(\ref{eq12}). Last column corresponds to the error defined by
(\ref{eq14}).}\label{Table 1}
\end{table}

The optimal value of convergence control parameter $h$ is further
calculated from equations (\ref{eqres1}-\ref{eqres2}) only for
$M=7$ as $h=1.30405$, which strongly indicates the reason of the
faster converge of the homotopy series near this value, see Table
\ref{Table 1}. Moreover, the uniform validity region of the
homotopy series solution (\ref{eq5}) can also be analytically
evaluated for the special value of $h=1$. In this particular case,
the terms in the homotopy series (\ref{eq5}) satisfy the ratio
\begin{eqnarray}\label{eq15}
  u_{n+1}(t)\over u_n(t) &=& {1\over 2}(1-e^{-2t}).
\end{eqnarray}Hence, regarding (\ref{eq15}), Theorem 1 assures
the convergence of the corresponding homotopy series (\ref{eq5})
for all values of $t$ valid in the interval $t>-{\ln 3\over 2}$.

\vspace{0.2in}{\bf Example 2.} Consider now the second-order
nonlinear differential equation
\begin{eqnarray}\label{eq16}
   2u''+u-u^2 &=& 0,\quad u(0) = 0,\quad u(\infty)=1,
\end{eqnarray}that governs the steady mixed convection flow past a plane
of arbitrary shape under the boundary layer and Darcy-Boussinesq
approximations \cite{Keller2001}. With the choices of $L={d^2\over
dt^2}-1$ and $u_0(t)=1-e^{-t}$, the homotopy (\ref{eq2}) generates
a homotopy series (\ref{eq5}) whose ratios of successive terms are
tabulated in Table \ref{Table 2} at several order of
approximations for some selected values of $h$. As observed from
the Table, $h=1.2$ generates divergent homotopy solution, and also
the convergence of the homotopy series to the exact solution takes
place at a considerably slow rate near $h=1$ as compared to the
smaller values of $h$.
\begin{table}[!htb]
\begin{center}
\begin{tabular}{|c|c|c|c|c|c|c|c|}
  \hline $h$ &$M=1$    & $M=10$  & $M=20$  & $M=30$  & $M=40$  & $M=50$  & $err$\\\hline
  1.2        & 0.42070 & 1.24479 & 1.31373 & 1.34407 & 1.35657 & 1.36525 & $8828.16$              \\
  1.0        & 0.24684 & 0.88919 & 0.94119 & 0.96001 & 0.96970 & 0.97513 & $3.444\times 10^{-4}$\\
  0.8        & 0.21484 & 0.63937 & 0.68278 & 0.69872 & 0.70700 & 0.71167 & $1.332\times 10^{-9}$\\
  0.6        & 0.36425 & 0.69932 & 0.74524 & 0.76240 & 0.77137 & 0.77643 & $1.753\times 10^{-7}$\\\hline
\end{tabular}
\end{center}
\caption{The ratio of successive terms in the homotopy series
corresponding to equation (\ref{eq16}). Last column corresponds to
the error defined by (\ref{eq14}).}\label{Table 2}
\end{table}

The optimal value of convergence control parameter $h$ is
calculated from equations (\ref{eqres1}-\ref{eqres2}) only for
$M=7$ as $h=0.73258$, which explains why the homotopy series
(\ref{eq5}) should converge faster near this value to the solution
of (\ref{eq16}), see Table \ref{Table 2}.

\vspace{0.2in}{\bf Example 3.} Consider now the nonlinear partial
differential Burger's equation
\begin{eqnarray}\label{eq17}
   u_t+uu_x &=& u_{xx},\quad u(x,0) = 2x,
\end{eqnarray}
that has been found to describe various kind of phenomena, such as
a mathematical model of turbulence and the approximate theory of
the flow through a shock wave traveling in a viscous fluid
\cite{Benton66a}. Equation (\ref{eq17}) admits an exact solution
given by
\begin{eqnarray}\label{eq18}
   u(x,t) &=& \frac{2x}{1+2t}.
\end{eqnarray}

To approximate the exact solution (\ref{eq18}), we choose the
auxiliary parameters as $u_0(x,t)=2x$ and $L={\partial \over
\partial t}$. Then, the homotopy (\ref{eq2}) turns out to be
\begin{eqnarray}\label{eq19}
   \left(1+\frac{1-p}{h p}\right)u_t(x,t,p)+u(x,t,p)u_x(x,t,p)-
   u_{xx}(x,t,p) = 0,\; u(x,0,p)=2x.
\end{eqnarray}
Equation (\ref{eq19}) produces the below homotopy series for the
solution of (\ref{eq17})
\begin{eqnarray}\label{eq19a}
   u(x,t) =
   2 x - 4 h t x + 4 h t (-1 + h + 2 h t) x -
 4 h t (-1 + h + 2 h t)^2 x \\\nonumber + 4 h t (-1 + h + 2 h t)^3 x -
 4 h t (-1 + h + 2 h t)^4 x+\cdots,
\end{eqnarray}whose convergence to the exact solution (\ref{eq18})
takes place for the values satisfying
$\lim_{n\to\infty}{|u_{n+1}|\over |u_n|}=|1 - h (1 + 2 t)|<1$.
Thus, in accordance with Theorem 1, this holds exactly for the
values of $-\frac{1}{2}<t<\frac{2-h}{2 h}$ together with $0<h<2$.
It is easy to demonstrate that the case $h=1$ corresponds to the
traditional Taylor series expansion of the solution (\ref{eq18}),
which is only valid in the region $-\frac{1}{2}<t<\frac{1}{2}$.

\vspace{0.2in}{\bf Example 4.} Consider now the following
fourth-order parabolic partial differential equation arising in
the study of the transverse vibrations of a uniform flexible beam
\cite{Gorman75}
\begin{eqnarray}\label{eq20}
   u_{tt}+({y+z\over 2\cos x}-1)u_{xxxx}+({z+x\over 2\cos y}-1)u_{yyyy}
   +({x+y\over 2\cos z}-1)u_{zzzz}&=& 0,\\\nonumber
   u(x,y,z,0) =-u_t(x,y,z,0)= x+y+z-(\cos x+\cos y+\cos z),
\end{eqnarray}
whose exact solution is given by
\begin{eqnarray}\label{eq21}
   u(x,t) &=& (x+y+z-\cos x-\cos y-\cos z)e^{-t}.
\end{eqnarray}
To approximate the exact solution (\ref{eq21}), if we choose the
auxiliary parameters respectively, $u_0(x,t)=(x+y+z-\cos x-\cos
y-\cos z)(1-t)$ and $L={\partial^2 \over \partial t^2}$, the
homotopy (\ref{eq2}) for $h=1$ then generates the subsequent
homotopy series
\begin{eqnarray}\label{eq22}
  u(x,t) &=& \sum_{n=0}^\infty u_n=\sum_{n=0}^\infty(x+y+z-\cos x-\cos y-\cos z)
  \left(\frac{t^{2n}}{(2n)!}-\frac{t^{2n+1}}{(2n+1)!}\right)
\end{eqnarray}which leads to $\lim_{n\to \infty} {|u_{n+1}|\over |u_n|}=0$.
So, the homotopy series (\ref{eq22}) converges to the exact
solution of (\ref{eq20}) for all $t$.

\vspace{0.2in}{\bf Example 5.} Let us consider the linear partial
differential equation
\begin{eqnarray}\label{eq23}
   u_t+u_x-2u_{xxt}&=&0,\quad u(x,0) =e^{-x},
\end{eqnarray}
whose exact separable solution is given by
\begin{eqnarray}\label{eq24}
   u(x,t) &=& e^{-x-t}.
\end{eqnarray}
To approximate the exact solution (\ref{eq24}), if we choose the
auxiliary parameters $u_0(x,t)=e^{-x}$ and $L={\partial \over
\partial t}$ respectively, then the homotopy (\ref{eq2}) turns out to be
\begin{eqnarray}\label{eq25}
   (1-p)u_t(x,t,p)+ p h(u_t(x,t,p)+u_x(x,t,p)-2
   u_{xxt}(x,t,p)) = 0,\; u(x,0,p)=e^{-x}.
\end{eqnarray}
When $h=1$, it appears that the homotopy series solution is
convergent only for $t\leq 0$, that is out of physical interest.
On the other hand, for $h=-1$, the successive ratio yields
$$|{u_{n+1}(t)\over u_n(t)}|={|t|\over n+1},$$ whose limit gives
rise to zero. This explains why the homotopy series solution
(\ref{eq5}) in this case represents the real physical solution
over $t\geq 0$ for $h=-1$, as also explained in \cite{Liang2009}.

\begin{figure}[!htb]
\begin{center}
\mbox{ \subfigure[]{
\includegraphics[width=6cm,height=5cm]{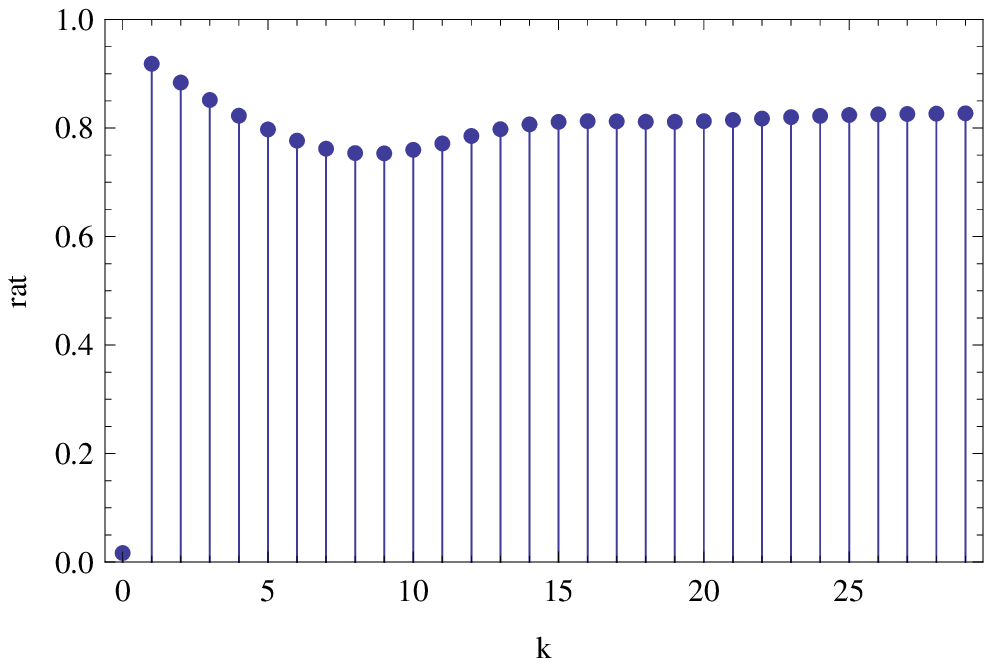}}
\subfigure[]{
\includegraphics[width=6cm,height=5cm]{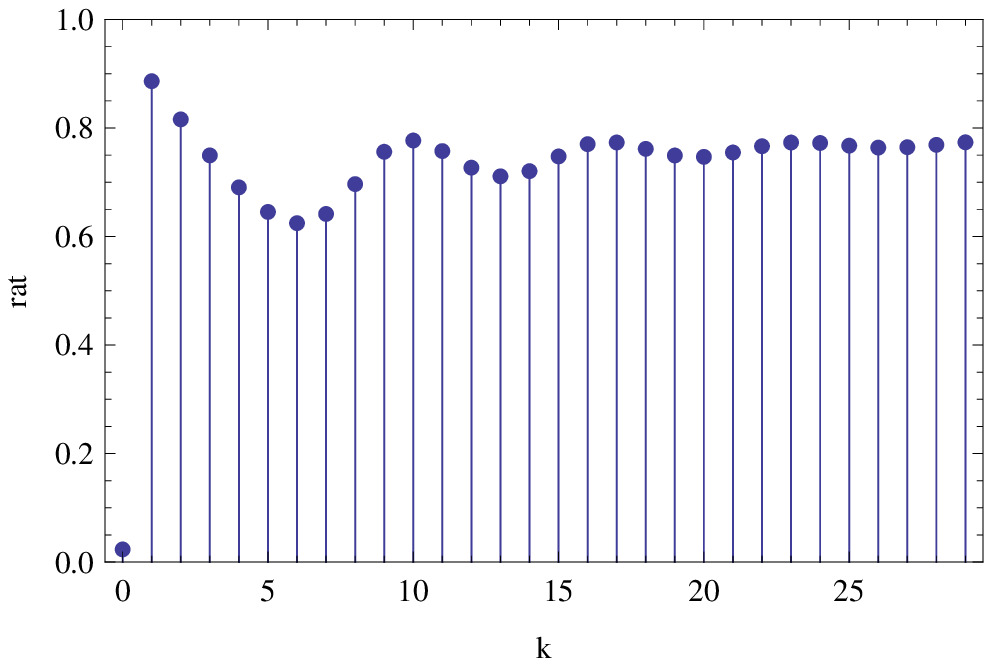}}}
\mbox{ \subfigure[]{
\includegraphics[width=6cm,height=5cm]{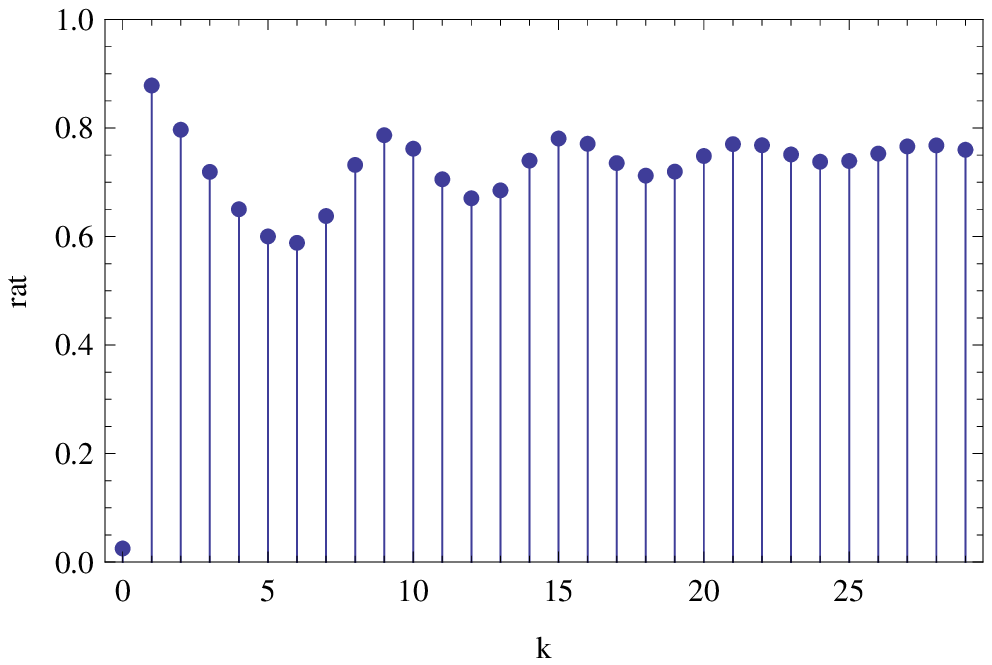}}
\subfigure[]{
\includegraphics[width=6cm,height=5cm]{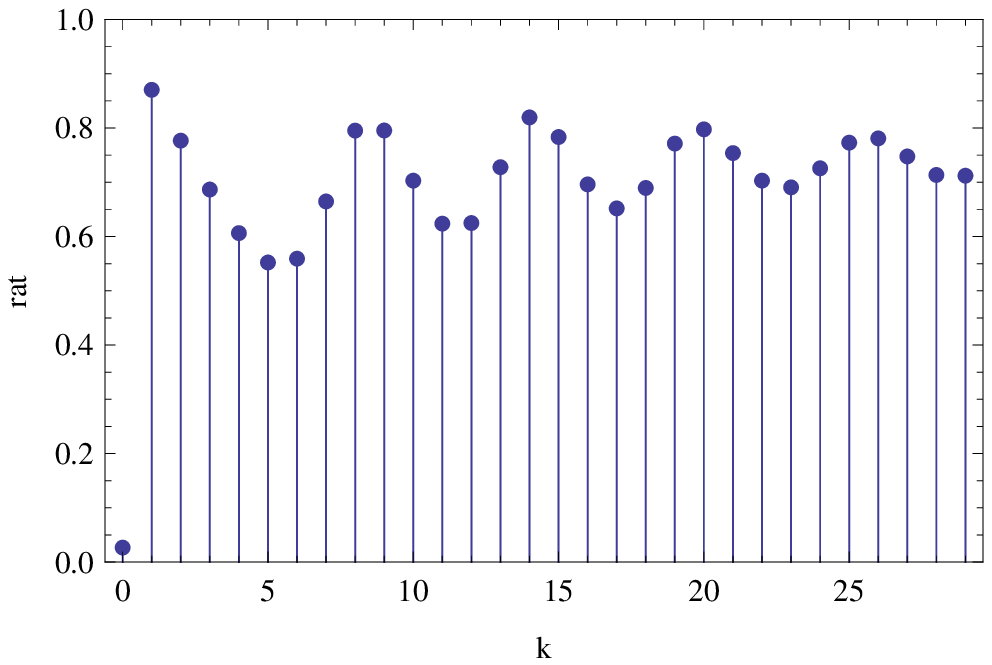}}}
\caption[Basic flow]{A list plot of the ratios $rat$ from the
theorem to reveal the convergence of the HAM solutions for the
Blasius equation (\ref{eq27}) for different choices of $h$. (a)
$h=-1$, (b) $h=-1.4$, (c) $h=-1.5$ and (d) $h=-1.6$.} \label{fig1}
\end{center}
\end{figure}

\vspace{0.2in}{\bf Example 6.} As a final example, we consider the
classical Blasius flat-plate flow problem of fluid mechanics
governed by the nonlinear initial-value of third-order
\cite{Liao99a}
\begin{eqnarray}\label{eq26}
   y'''+{yy''\over 2} &=& 0,\quad y(0)=y'(0)=y'(\infty)-1=0,
   \quad \eta \in [0,\infty).
\end{eqnarray}With the transformations $$y={u\over \lambda},\quad t=\lambda \eta,$$
where $\lambda$ is a scaling parameter taken here as 4, system
(\ref{eq26}) is converted into
\begin{eqnarray}\label{eq27}
   u'''+{uu''\over 2\lambda^2} &=& 0,\quad u(0)=u'(0)=u'(\infty)-1=0,
   \quad t\in [0,\infty).
\end{eqnarray}
Taking $L={d^3\over dt^3}+{d^2\over dt^2}$ with
$u_0(t)=-1+t+e^{-t}$, the values for the convergence control
parameters $h$ from the homotopy (\ref{eq2}) are shown in Table
\ref{Table 3} along with the corresponding residual errors. It is
seen that the optimal value for the physical problem considered is
$h=-3/2$. Figures \ref{fig1}(a-d) demonstrate the ratios
(\ref{rat}) in Theorem 1 which were computed from the homotopy
series (\ref{eq5}) for the values of $h$ given in Table \ref{Table
3}. The convergence is assured further by the values of the ratios
less than unity, suggesting that limit tends to 0.83, 0.77, 0.76
and 0.78, for the values of $h$ shown.
\begin{table}[!htb]
\begin{center}
\begin{tabular}{|c|c|c|c|c|c|}
  \hline $h$ & $M=1$    & $M=10$                & $M=20$               & $M=30$               \\\hline
  1.6        & 1.258621 & $1.201\times 10^{-3}$  & $2.463\times 10^{-7}$ & $1.032\times 10^{-6}$ \\
  1.5        & 1.027450 & $1.372\times 10^{-4}$  & $3.602\times 10^{-7}$ & $2.573\times 10^{-9}$ \\
  1.4        & 0.820771 & $1.399\times 10^{-4}$  & $7.621\times 10^{-7}$ & $5.454\times 10^{-9}$ \\
  1.0        & 0.238796 & $7.621\times 10^{-4}$  & $1.511\times 10^{-5}$ & $4.540\times 10^{-7}$ \\\hline
\end{tabular}
\end{center}
\caption{The square residual errors $Res(h)$ for the homotopy
series solutions corresponding to equation (\ref{eq27}) at several
values of convergence control parameters $h$.}\label{Table 3}
\end{table}

\section{Concluding remarks}
\label{conclusions} In this paper, the homotopy analysis method
has been analyzed with an aim to investigate the conditions which
result in the convergence of the generated homotopy solutions of
the nonlinear ordinary and partial differential equations. The
theorems outlined in the paper have proved that if specific values
are assigned to the auxiliary parameters in the homotopy analysis
method, then the approximate homotopy results successfully
converge to the exact solution. An optimal value approach for the
convergence control parameter has also been given. Examples have
been provided to verify the theory. Via the theorems provided
here, not only the question of the convergence of the homotopy
series is answered, but also the region of validity of the space
variable ensuring the convergence is determined. The traditional
flat-plate boundary layer flow problem has been finally treated by
the convergence theorem.
\bibliographystyle{unsrt}
\bibliography{mybib}
\end{document}